\newcommand{\ergps}{erg\thinspace s$^{-1}$}
\newcommand{\ergpspsqcm}{erg\thinspace cm$^{-2}$\thinspace s$^{-1}$}
\newcommand{\psqcm}{cm$^{-2}$}

\documentclass[final,3p,times,twocolumn]{elsarticle}

\usepackage{graphicx}

\usepackage{amssymb}

\journal{New Astronomy}

\begin{document}

\begin{frontmatter}



\title{The gravitationally lensed, luminous infrared galaxy IRAS
  F10214+4724 observed with XMM-Newton}


\author[1]{K. Iwasawa}
\address[1]{Osservatorio Astronomico di Bologna, Via Ranzani 1, 40127 Bologna, Italy}
\author[2]{C. Vignali}
\address[2]{Dipartimento di Astronomia, Universt\`a di Bologna, Via Ranzani 1, 40127 Bologna, Italy}
\author[3]{A.S. Evans}
\address[3]{Department of Astronomy, University of Virginia, 530 McCormick Road, Charlottesville, VA 22904 and NRAO, 520 Edgemont Road, Charlottesville, VA 22903-2475, USA}
\author[4]{D.B. Sanders}
\address[4]{Institute for Astronomy, 2680 Woodlawn Drive, Honolulu, Hawaii 96822-1839, USA}
\author[5]{N. Trentham}
\address[5]{Institute of Astronomy, University of Cambridge, Madingley Road, Cambridge CB3 0HA, United Kingdom}

\begin{abstract}
We report on a short XMM-Newton observation of the
gravitationally-lensed, luminous infrared galaxy IRAS F10214+4724 at
$z=2.3$. A faint X-ray source is detected at $4 \sigma$. The observed
0.5-2 keV (1.7-6.6 keV in the rest-frame) flux is $1.3\times 10^{-15}$ \ergpspsqcm and the spectral
slope in the rest-frame 1-10 keV band is $\Gamma\sim 2$. These
results agree with those obtained from the Chandra X-ray
Observatory, given the large uncertainties in both measurements. While
possible evidence for excess emission above 5 keV is seen, we suspect
this excess might be either spurious or not related to the infrared
galaxy.
\end{abstract}


\end{frontmatter}


\section{Introduction}
\label{}

IRAS F10214+4724 is an IRAS source identified with an emission-line
galaxy at $z=2.286$ (e.g., Rowan-Robinson et al 1991; Soifer et al
1991). The extremely large infrared luminosity $\sim 10^{14}$
$L_{\odot}$ is partly due to the amplification of gravitational
lensing by an intervening galaxy or a group of galaxies at $z\sim 0.9$
(Broadhurst \& Lehar 1995; Goodrich et al 1996; Lacy, Rawlings \&
Serjeant 1998). The amplification factor appears to have wavelength
dependency and estimated to be in the range of 10-100 (Broadhurst \&
Lehar 1995; Downes et al 1995; Eisenhardt et al 1996; Evans et al
1999; Graham \& Liu 1995; Serjeant et al 1995; Trentham 1995; ).

The presence of an active nucleus in this galaxy was already clear
from the optical emission-line spectrum of Seyfert 2 (Elston et al
1994; Soifer et al 1995) and the detection of polarised broad emission
(Goodrich et al 1996) suggested that the nucleus is obscured from the
direct view. The Spitzer IRS spectrum, however, shows a silicate
feature in emission, in contrast to other obscured objects, of which
the mid-infrared spectra generally show the silicate features in
absorption (Teplitz et al 2006). The PAH features are weak or absent,
indicating the mid-infrared light may be dominated by AGN (discussion
on the PAH features in ULIRGs and PG quasar host galaxies can be found
in Schweitzer et al 2006; Shi et al 2007; Armus et al 2007).

A large resevoir of molecular gas ($\sim 10^{11} M_{\odot}$, Brown \&
Van den Bout 1992; Radford, Brown \& Van den Bout 1993; Radford et al
1996; Scoville et al 1995; Wagg et al 2006) has been inferred from CO
observations. Strong star formation could, in principle, take place
there, and the large far-infrared excess due to cold dust may
originate from this. If this region has a large optical depth,
mid-infrared emission from the starburst could be suppressed. In
modelling the infrared energy distribution, Efstathiou (2006)
attributes a deeply buried starburst to the far-infrared to
sub-millimetre emission while the Seyfert 2 nucleus to the emission at
shorter wavelengths (also offers an explanation for the silicate
emission).

A 20-ks Chandra observation has detected 14 counts from a faint X-ray
source associated with this object (Alexander et al 2005). We have
also obtained the position of the Chandra X-ray source using the
latest calibration: (RA, Dec.)$_{\rm J2000} = (10^{\rm h}24^{\rm
  m}34^{\rm s}.57, +47^{\circ}09^{\prime}09^{\prime\prime}.9$), which
is 0.4 arcsec to the north of the position reported previously but in
a good agreement with the position of the millimetre continuum source
measured with IRAM (beam size of $\sim 1$ arcsec, Ao et al 2008). With
the positional uncertainty within 0.5 arcsec, the association of
the X-ray source with the infrared galaxy is secure, as discussed in
Alexander et al (2005). Here, we report an update on the X-ray source,
obtained from a short XMM-Newton observation\footnote{Since this XMM
  observation was approved in the same period as for the Chandra
  observation, we had no prior knowledge of the X-ray brightness of
  this object, apart from the upper limits from ROSAT (Lawrence et al
  1994) and ASCA (Iwasawa 2001).}. The cosmological parameters used
for calculating the source luminosity are the Hubble constant $H_0 =
70$ km s$^{-1}$ Mpc$^{-1}$, $\Omega_{\rm m}=0.27$, and
$\Omega_{\lambda}=0.73$.

\section{XMM-Newton Observation}

IRAS F10214+4724 was observed with XMM-Newton on 2004 October 22-23
(Observation ID: 0201040101). The observation with the EPIC (one pn
and two MOS) cameras, which are used for the analysis presented in
this paper, was performed in the Imaging Full-Window mode with Thin
Filter for the pn and Medium Filter for the MOS cameras. A significant
fraction of the observation with the nominal 50 ks duration was lost
by strong flares of the particle background. The useful exposure times
for the scientific analysis, taken from periods of quiescent
background, are 20.2 ks for the pn and 33.8 ks for the MOS data. The
source counts were collected from a circular region with a radius of
12.5 arcsec, centred on the optical position, and the background data
were taken locally from a source-free region on the same detector. The
data reduction and analysis were carried out using the software
packages of SAS 8.0.1 and HEASoft 6.6.1.

\section{Results}

\begin{figure}
\centerline{\includegraphics[width=0.37\textwidth,angle=0]{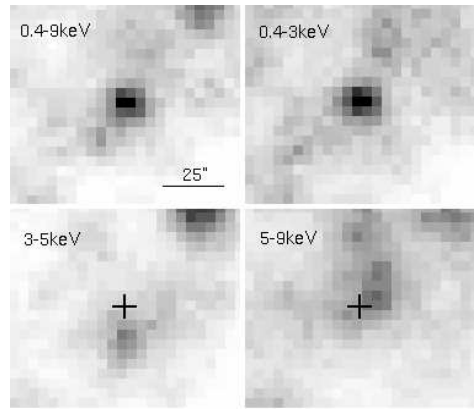}}
\caption{The XMM-Newton images of IRAS F10214+4724. The images are
  made by adding the EPIC pn and MOS data, and then smoothed using a
  circular top-hat filter. The filter kernel is of an adaptive size so
  that at least 25 counts (for the 0.4-9 keV image) / 16 counts (for
  the others) are contained under the filter. The energy band where
  the photons are collected is indicated in the figure: 0.4-9 keV
  (1.3-30 keV), 0.4-3 keV (1.3-9.9 keV), 3-5 keV (9.9-17 keV) and 5-9
  keV (17-30 keV), where the rest-frame energies are in (). The cross
  shows the position of the soft X-ray peak, which coincides with the
  optical position of the galaxy. A bar in the 0.4-9 keV image
  indicates the angular scale of 25 arcsec.}
\end{figure}

The background-corrected counts in the 0.4-9 keV band are $14.2\pm
6.3$ from the pn and $21.7\pm 6.5$ from the two MOS
cameras. Expected source counts from the count rate detected with
Chandra, assuming the power-law slope of $\Gamma = 1.6$ (the effective
Chandra slope based on the hardness ratio, Alexander et al 2005)
modified by the Galactic absorption $N_{\rm H}=1.25\times 10^{20}$ cm$^{-2}$
(Kalberla et al 2005), are 24 for the pn and 22 for the two MOS
added at the respective exposure times. The MOS counts are in good
agreements. The pn counts are $\sim 40$ per cent smaller than
expected, but given the uncertainty of the pn counts and that the
Chandra counts are only 14 with an approximately 30 per cent
error, they are consistent within the uncertainties. 

The X-ray images in the 0.4-9 keV band and the three sub-divided bands
(0.4-3 keV, 3-5 keV, and 5-9 keV), obtained by adding the three EPIC
cameras, are shown in Fig. 1. A source is detected at the optical
position of IRAS F10214+4724. The detection is at $4\sigma $ in the
soft (0.4-3 keV) band, which corresponds to the rest-frame 1.3-10
keV. There is no detection in the 3-5 keV, while the 5-9 keV band
shows a $\sim 2\sigma $ excess at the position of the galaxy. This
possible 5-9 keV excess is seen both in the pn and MOS data. However,
on inspecting the 5-9 keV image, this excess could be due to a
contamination from a nearby source or more likely, just noise due to
background uncertainty, as no clear source peak can be seen at the
expected position as seen in the soft band (Fig. 1). No other X-ray
source is seen in the Chandra image within the XMM-Newton aperture.

If this were a real detection, it could be due to a strongly
absorbed component appearing above 15 keV. This energy of absorption
cut-off represents a column density of $N_{\rm H}\sim 10^{25}$
\psqcm. However, at such a high column density, the corresponding
large Thomson depth means that multiple Compton down-scattering would
suppress high-energy photons and no large excess at high energies
would be expected.  Therefore, together with the small significance of
the excess counts, we do not consider the 5-9 keV excess to be a
reliable detection for the object of our interest. This also applies
to the 5-9 keV data point of the spectrum (Fig. 2).

\begin{figure}
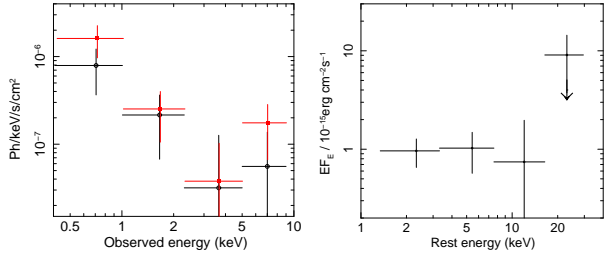


\hbox{\centerline{\includegraphics[width=0.2\textwidth,angle=270]{f2a2.ps}\hspace{0mm}
\includegraphics[width=0.2\textwidth,angle=270]{f2b.ps}}}
\caption{The XMM-Newton EPIC spectrum of IRAS F10214+4724. The left
  panel shows the photon spectrum as observed (pn: open circles in
  black; MOS: filled squares in red), while the right panel shows the
  flux spectrum with the rest-frame energies. As discussed in the
  text, the highest energy point is likely to be spurious.}
\end{figure}

Given the limited source counts, only a coarse spectrum with four
spectral channels is constructed for a rough estimate of the spectral
shape, by combining the pn and MOS data. The pn and MOS spectra were
corrected for the background and the detector response curve
individually, and verified to be in a good agreement in their shapes
(Fig. 2, left). As reported above for the count rate, the
normalisation of the pn spectrum is $\sim 40$ per cent lower than the
MOS spectrum. The four-channel flux spectrum (equivalent to $\nu
F_{\nu}$) as a function of rest-frame energies is shown in Fig. 2
(right). When the unreliable 5-9 keV data point is excluded, fitting
with a power-law modified by the Galactic absorption gives $\Gamma =
2.0^{+0.6}_{-0.5}$ (a fit of the individual pn and MOS spectra gives
the same result). No clear evidence for a hard X-ray excess due to an
absorbed AGN component can be found. The observed flux estimated from
the power-law model is $1.3\times 10^{-15}$ \ergpspsqcm\ in the 0.5-2
keV band, and $0.9\times 10^{-15}$ \ergpspsqcm\ in the 2-5 keV
band. Note that these fluxes are the mean of the pn and MOS data. The
rest-frame 2-10 keV luminosity of the intrinsic X-ray source is
estimated to be $f_l^{-1}6.4\times 10^{43}$ \ergps, where $f_l$ is the
lensing amplification factor (see Introduction).

\section{Discussion}

The XMM-Newton data give slightly lower flux and steeper spectral
slope than those from the Chandra observation, but they agree within
the uncertainty. The apparent spectral shape with $\Gamma\sim 2$ in
the rest 1.3-10 keV band is consistent with an unobscured AGN, which
is not in agreement with the hidden AGN suggested e.g., by the
spectropolarimetry (Goodrich et al 1996), unless the X-ray absorption
is only moderate with $N_{\rm H}\sim 10^{22}$ cm$^{-2}$ (e.g.,
Severgnini et al 2006), in which case the absorption cut-off would
fall below the bandpass in this high redshift object. There is also a
possibility that extended soft X-ray emission, e.g., from a starburst,
could mask the absorbed contniuum of the AGN to mimic the steep
spectrum in a low resolution spectrum. Given the small number of
detected counts, no meaningful constraint on the Fe K line, hence on a
possible presence of heavily obscured (Compton-thick) active nucleus,
can be obtained. 

In a strongly gravitationally lensed system, the amplification due to
the lens can vary significantly between emission regions, e.g.,
infrared emitting dust region versus an X-ray emitting nucleus. This
effect results in a distortion of the spectral energy distribution
over a wide wavelength range to an observer on the Earth. This means
that a direct comparison of the X-ray luminosity and the infrared
luminosity (which is the bulk of the bolometric luminosity) may not
provide a reliable measure of the AGN contribution to the bolometric
luminosity. In IRAS F10214+4724, although the detail of the lensing
configuration is still highly model dependent, the extended galaxy
emission is considered to lie closer to the lensing caustic than the
nuclear emission, resulting in the stellar and extended dust emission
being more amplified than the active nuclues (Evans et al 1999, but
see also Teplitz et al 2006). If the observed X-ray emission is assumed to
be due entirely to an unobscured or moderately obscured AGN, a
straight conversion from the 2-10 keV luminoisity to the bolometric
luminosity, even with a large bolometric correction $\sim 100$,
appropriate for luminous quasars (e.g., Marconi et al 2006), would
fall $\sim 2$ orders of magnitude below the far infrared
luminosity. This X-ray quietness may point to an unfavourable lensing
configuration for the innermost part of the accretion disk where the
primary X-rays are produced, or alternatively, the faint X-ray
emission in IRAS F10214+4724 being only reflected light from a heavily
obscured AGN (see Alexander et al 2005 for details).

An assessment of the AGN contribution to the bolometric luminosity and
the lensing amplification factor could only be possible once the
origin of the observed X-ray emission has been identified either with,
e.g., direct light from the active nucleus, reflected light of a
Compton-thick AGN, or thermal emission from a starburst. The X-ray
amplification factor should differ accordingly, as the physical scale of the
respective origins is vastly different. The primary X-ray emitting
region would be at $5\times 10^{-5}(M_{\rm BH}/10^8\thinspace
M_{\odot})(R_{\rm X}/10\thinspace r_{\rm g})$ pc, where $M_{\rm BH}$
is the black hole mass, $R_{\rm X}$ is the size of the X-ray
production region in unit of the gravitational radius, $r_{\rm g}\equiv
GM_{\rm BH}/c^2$. X-ray reflection from a Compton thick AGN is
expected to occur at the inner surface of the torus at $\leq 1$ pc, if
its optical depth there is large enough, or possibly a more extended region
of dense molecular gas where the CO observations probed. Starburst
related emission would come from a range of radii, depending on the
energy, but primarily from a similar region to where the
FIR-submillimetre emission arises. A sufficiently deep X-ray
observation would be necessary to identify the origin of the X-ray
emission. 

\bigskip

We thank D.M. Alexander for useful discussion. This paper is based on
observations obtained with XMM-Newton, an ESA science mission with
instruments and contributions directly funded by ESA Member States and
NASA.

\end{document}